# Returning the Favor: What Wireless Networking Can Offer to AI and Edge Learning

Sameh Sorour, Umair Mohammad, Amr Abutuleb, and Hossam Hassanein

*Abstract* — **Machine learning (ML) and artificial intelligence (AI) have recently made a significant impact on improving the operations of wireless networks and establishing intelligence at the edge. In return, rare efforts were made to explore how adapting, optimizing, and arranging wireless networks can contribute to implementing ML/AI at the edge. This article aims to address this void by setting a vision on how wireless networking researchers can leverage their expertise to return the favor to edge learning. It will review the enabling technologies, summarize the inaugural works on this path, and shed light on different directions to establish a comprehensive framework for mobile edge learning (MEL).**

## INTRODUCTION

In recent years, ML and AI have flourished as key contributors to many research fields. Wireless networking is certainly one of these fields that has significantly benefited from ML/AI on all levels, from radio resource management, to medium access, network organization, routing, caching, service discovery, wireless-based positioning, crowdsensing, and mobile edge computing (MEC). Researchers in these areas have extensively resorted to ML/AI to develop efficient solutions to intractable problems and simpler implementations for hard-to-implement ones. Contrarily, very limited works in the wireless networking realm aimed to facilitate the operation of ML/AI. A question on whether this is possible has always inspired wireless networking experts.

The answer to this question was not revealed until very recently, when the rapid adoption of the Internet of Things has resulted in the generation of exploding amounts of data at the network edge. The massive sizes, distributed nature, and – in many applications – time-criticality of analyzing this data, clearly prohibit their centralized processing. Indeed, transferring these data volumes to cloud servers across entire cities via multiple nodes/links will take too long, cost too much, and increase security and privacy risks [1]. Forecasts predict that the application of ML/AI on 90% of this data must be done at edge servers or more preferably edge devices. Since most edge devices are wireless/mobile (e.g., wireless sensors, smartphones, laptops, drones, connected vehicles) and resource-constrained, implementing ML/AI on networks of such devices will be fundamentally contribute to this trend, thus giving birth to the new notion of MEL.

MEL can be thought of as the synergistic evolution of two paradigms undergoing significant advancements lately, namely distributed learning (DL) and MEC. DL has recently attracted much attention within the ML realm [2] to enable the training of global learning models using parallelly-running models on a set of edge nodes. DL has been mostly studied in wired and homogeneous environments, and its extension to wireless/mobile domains was not explored until very recently. Yet, the most important trait of wireless networks, namely their nodes' inherent heterogeneities and constraints, was commonly overlooked. On the other hand, MEC has exhibited outstanding potentials in handling independent computing jobs of resource-constrained and heterogeneous wireless/mobile devices, by optimizing their offloading to edge servers [3]. Fewer recent works considered more general strategies that involved offloading among peer devices. Yet, the execution of DL jobs was never explored in MEC settings.

The implementation of such jobs on wireless/mobile devices thus calls for wireless networking and computing experts to design a revolutionary framework for MEL. This framework must resolve several integration problems between DL and MEC to efficiently execute MEL jobs. These problems can be classified under two major directions, namely task/network optimization and network formation. The former direction involves developing mechanisms to jointly allocate resources and tasks to devices to fulfill MEL's learning requirements given the devices' heterogeneous networking, computing, and energy constraints. It must also leverage the mobility of these devices to enhance learning outcomes and/or durations. Developing mitigation strategies to service/connection loss from one/multiple devices, due to channel impairments, user behavior, or mobility, is another requirement for MEL's success.

The latter direction will focus on adapting the network structure to the number, types, and complexity of the MEL jobs. It must optimally partition the available devices into different clusters to perform multiple simultaneous jobs. It should also derive optimal routing and model aggregation policies in multi-hop network settings. Scenarios involving complex learning models may additionally require structuring the available limited-capacity devices into cascaded chains and splitting these models among them.

This article aims to streamline the required research efforts that can collectively materialize a robust framework for MEL and highlight how networking experts can contribute

- *S. Sorour, Amr Abutuleb, and H. Hassanein are with the School of Computing and Dept. of Electrical and Computer Eng., Queen's University, Kingston, ON, Canada. E-mail: {sameh.sorour, amr.abutuleb, hassanh}@queensu.ca.*
- *Umair Mohammad is with the Dept. of Electrical and Computer Eng., University of Idaho, Moscow, ID, USA. E-mail: moha2139@vandals.uidaho.edu.*

to these efforts. The reminder of this article will thus proceed as follows. The next two sections will elucidate the key enablers of MEL, namely DL and MEC. The following section will illustrate how these enablers can synergistically give birth to MEL and list the challenges facing this endeavor. The two subsequent sections will shed light on how networking experts can leverage their expertise to contribute to the two aforementioned directions towards establishing a comprehensive framework for MEL.

## ENABLING TECHNOLOGIES

### Distributed Learning

DL has recently drawn significant attention within the ML realm motivated by two practical scenarios, namely learning federation and parallelization. In federated learning (FL), a node (a.k.a. orchestrator) wants to learn from separately collected/stored datasets at multiple nodes (a.k.a. learners) without transferring these datasets to itself. This setting becomes of high interest when these datasets are either too big to transfer given the available bandwidth or too private to share with the orchestrator. FL operates by executing "global learning cycles", each consisting of three steps:

1. The orchestrator distributes the learning model and global parameters to the learners.
2. Each learner runs a number of local training cycles on the received model (a.k.a. local model) using a subset or all its dataset.
3. The orchestrator collects and aggregates the learners' local parameters to update the global parameters.

Fig. 1 (Left) shows the steps of one FL global cycle. The orchestrator typically executes multiple cycles until a desired learning quality level (LQL) or time deadline is reached.

Conversely, parallelized learning (PL) involves a node wanting to perform ML on its collected/stored dataset while lacking the resources (e.g., CPU power, energy) or time to perform it locally. It thus parallelizes this job on multiple learners (which can range from different cores of a computing cluster to standalone nodes) and acts as orchestrator by running global learning cycles as in FL. A major difference in PL is that the orchestrator must distribute new samples of its dataset to learners in each global cycle to train their local models with. Fig. 1 (Right) illustrates the steps of one PL global cycle.

The quality and convergence rate of DL are significantly impacted not only by the numbers of local training cycles performed by learners per global cycle, but also by the variation in these numbers between different learners (a.k.a. staleness). If learning jobs are given enough time/resources, minimizing this variation was reported to yield the best performance [4]. Thus, the best setting is *synchronous* DL, in which learners perform an equal number of local cycles. Yet, unlike homogeneous infrastructure-based learners, this synchronicity may be limiting and non-scalable among highly heterogenous wireless/mobile ones, as it will be constrained by the least-capable device(s). This will clearly restrict the potentials of more capable devices in achieving higher LQL and/or convergence rate. The only solution in such instances is to resort to *asynchronous* DL [4,5], which allows for some minimized/controlled staleness if full synchronicity cannot be achieved or would degrade the performance.

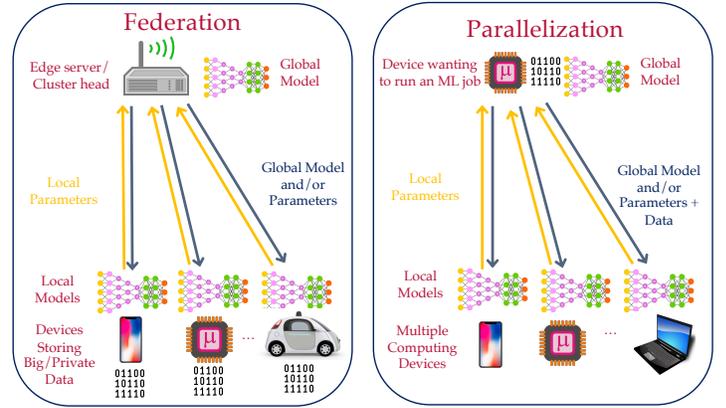

Fig. 1: Illustration of an example setup of FL (left) and PL (right) in wireless edge environments.

### Mobile Edge Computing

MEC is an established paradigm that regulates the offloading of typically independent computational tasks from wireless/mobile devices to installed computing servers in their vicinity (e.g., at base-stations, roadside units). This offloading involves the transmission of tasks' descriptions and data to edge servers, their execution by these servers, and the delivery of results to devices. Devices' savings on completion time and/or consumed energy are the usual sought benefits when making task offloading decisions [3].

Different MEC problems have been investigated in the literature. Offloading decisions to one or multiple edge servers were optimized both separately and jointly with the network resources. Device mobility was also leveraged to achieve better offloading outcomes. On the server side, multi-level architectures, in which tasks can be offloaded to edge or cloud servers, were suggested. Other studies explored joint offloading, resource, and trajectory optimizations for drone-mounted MEC servers.

Only recently were unoccupied peer devices involved as (additional) resources to offload tasks to [6]. This option unleashed a new dimension of edge cooperation, where server(s) and devices become a networked resource pool to which tasks can be offloaded, shuffled, and/or parallelized to achieve local and/or global benefits. Yet, no significant studies explored the potentials of such networked MEC pools in handling correlated computing/DL tasks nor optimized these pools' resources, procedures, and structures to fulfill the requirements of such inter-dependent tasks.

## TOWARDS A FRAMEWORK FOR MEL

### Foundation

It can be inferred from the previous discussions that DL implementations on networked pools of MEC resources can lay the foundation for the desired MEL framework. From DL's perspective, the constrained settings in mobile edge environments are the upmost manifestation of the need for FL and PL. Indeed, wireless/mobile devices may not be able to transfer their generated/collected data to edge servers due to bandwidth limitations and/or data privacy. Many of these devices are also computationally limited and may need to parallelize their learning jobs on peer devices for faster processing and/or lower energy consumption. The major differences between conventional DL and MEL are the high heterogeneity and diverse constraints among the learners. These differences have direct implications on the three steps of DL's global cycles:

- The learners' limitations and heterogeneities in channel qualities and energy budgets will cause significant reductions and variations in the orchestrator's speeds to deliver the learning model/parameters (+data for PL) to these learners and collect the local parameters from them.

- The heterogeneities in learners' computational capacities and battery constraints will significantly impact the amount of data and number of local cycles each learner can execute within each global cycle.

The implications of these heterogeneities and constraints are thus game-changing, and their impact on optimizing task and resource allocation to learners need to be thoroughly investigated.

From MEC's perspective, the implementation of DL in mobile edge environments can be viewed as a novel sophisticated MEC paradigm, in which one or multiple learning jobs are sliced into highly correlated tasks and assigned/offloaded to a pool of highly heterogeneous and constrained nodes. Thus, the operation of these nodes requires high levels of coordination and joint optimizations of task/resource assignments to achieve quality metrics not only on the networking and computing sides but also on the outcomes of the learning jobs.

## Challenges

Given the above description, several challenges need to be addressed to collectively build a new MEL framework:

*Allocating Tasks and Resources*: In MEL, the learner's task size is defined by the number of data samples it should train its local model with in one global cycle or the entire learning process. The orchestrator typically determines these task sizes. It then either instructs the learners to select as many data samples as determined from their own stored data in FL or sends the determined number of data samples to them in PL. Due to resource and channel heterogeneities, allocating equal or arbitrary (e.g., random sizes, all stored data in FL) task sizes to all learners may neither be optimal nor even feasible given the time constraints on MEL jobs. Adapting task sizes to learners' resources and channel qualities is thus needed to maximize the number of learning cycles executed within each global cycle. Furthermore, the selection of learners' physical resources (e.g. transmit power, CPU shares) and procedures (e.g., channel assignment) can be jointly optimized with task allocations to yield the upmost performance.

*Mitigating Physical Uncertainties*: Wireless/mobile devices may exhibit connection degradation or losses due to wireless channels impairments (e.g., fading, interference, collisions, etc.) and/or device mobility (e.g., handover failure, exiting coverage range). Additionally, the allocated computing power by any device to the DL job may suddenly change due to many factors (e.g., internal interrupts, processor freezes, variations of load by owners). Proactive and reactive measures should thus be implemented to mitigate the risks of such uncertainties and make sure the orchestrator receives all parameters at the desired time with the best quality.

*Leveraging Mobility*: In some scenarios, the learners' locations relative to the orchestrator may not be optimum. If these learners' mobility is controllable (e.g., autonomous vehicles, mobile sensors), adjusting their positions may significantly improve MEL's performance. In settings involving mobile orchestrators (e.g., drone-mounted edge server), their trajectories can be planned given the positions and capabilities of the learners or jointly optimized with their mobility.

*Handling Multiple Jobs*: Many scenarios may necessitate the simultaneous execution of multiple independent or correlated learning jobs in the same edge environment. Strategies must thus be developed to partition this environment's devices into different sub-networks, based on their heterogeneous resources, diverse connection qualities to the orchestrator(s) and/or their stored data (in FL), to perform these simultaneous MEL jobs.

*Handling Complex Learning Models*: Learning models are getting increasingly complicated. Some problems even require the concatenation of multiple learning models to achieve the desired performance (e.g., concatenating a convolutional neural network with a recurrent neural network for video analytics). The implementation of such models on some or all devices may not be feasible due to their limited processing capabilities, insufficient memories, battery limitations, and/or owners' settings. Innovating optimized approaches to have these learners host stages of these complex models and complement one another's work, given their networking, computing, and energy credentials, may be the only resort in such scenarios. Having highlighted these various challenges, the next two sections will shed light on how the expertise in wireless networking and mobile computing can be leveraged to address these challenges and serve the purpose of establishing an adaptive, dynamic, and scalable MEL framework.

*Operating in Multi-Hop Settings*: Multi-hop networking is very typical in a broad range of wireless environments (e.g., wireless sensor networks, vehicular networks). Performing MEL jobs in such multi-hop settings will thus necessitate the derivation of optimal routing and in-route parameter aggregation approaches.

## TASK/NETWORK OPTIMIZATIONS

### Task Adaptation

As mentioned earlier, the simplest and most direct type of optimization within a network of MEL learners is to adapt the allocated task sizes to the physical resources (e.g., transmit power, CPU cycles) of these learners. This adaptation can be performed to optimize various metrics that improve the LQL by the end of each global cycle or the entire training process. Examples of such metrics are minimizing the learning loss function, maximizing the number of synchronous or asynchronous local cycles per global cycle (with bounds on staleness in the asynchronous case), or minimizing staleness among different learners.

It can be easily inferred that each of the steps of a global cycle is significantly impacted by learners' resources. Indeed, the times taken by Steps 1 and 3 for each learner are strongly governed by the transmit power of the orchestrator and this learner, respectively, as well as their channel bandwidth and quality. In PL, Step 1's duration is also impacted by the size of the data samples (i.e., task size) sent from the orchestrator to this learner. Furthermore, the time needed to complete Step 2 by each learner is governed by both the task size and this learner's available CPU speed. All three steps are also tied by each device's battery level and its imposed limits on the energy it can consume to contribute to the MEL job. All these physical resources, qualities, and constraints being strongly heterogenous in mobile edge environments, the task size allocated to each learner should thus be carefully chosen so as to optimize any of the aforementioned metrics within the allotted time and/or allowed energy budget for each global cycle. If time and/or energy constraints are imposed on the entire training process, the number of global cycles, local cycles, and task sizes can all be jointly optimized to achieve the desired target while satisfying these constraints.

As simple as these adaptations are, it has been shown that they can yield significant gains [7-9]. Fig. 2 illustrates the impact of adapting task allocations to network resources compared to when not adapting them. For the well-known MNIST dataset, the figure depicts the progression of the testing accuracy of a trained 5-layer neural network on 20 learners after each 12-seconds-long global cycle. It can be seen from the figure that a reduction of 33% and 50% are achieved in the number of global cycles that are needed to reach a testing accuracy of 94% and 97%, respectively.

### Joint Task and Resource Optimization

The next level of network adaptation to MEL is the one that jointly optimizes task allocation with the physical resources (e.g., transmit powers, channel bandwidth, and CPU speed) and procedures (e.g., channel assignment, medium access) of the different learners to achieve even further gains. Some recent works have started to investigate physical resource optimization for FL [10-12]. In addition, few works have explored the option of data re-distribution among devices to minimize the FL training time [13]. This typically adds an extra step to the global cycle, in which devices exchange data to reach its best distribution among them. The optimization in these works is thus concerned with determining the physical resources and exchanged data sizes to minimize the overall FL training time. However, all these works ignored task size optimization and forced the learners to perform the training using all their stored data. This may be acceptable when the total FL training time is not constrained but may clearly fail when it is, which is very typical in MEL jobs. In addition, none of these works have considered the PL side of MEL, which is more impacted by these wireless networking considerations as it involves the transmission of data samples in each global cycle. We have attempted to fill the above gaps by jointly optimizing task sizes and physical resources for both FL and PL. For the same setting of Fig. 2, Fig. 3 depicts the gains of this joint optimization as opposed to only optimizing either task allocation or physical resources. For example, 50% and 57% reduction in the number of global cycles to reach 98% testing accuracy are achieved compared to optimizing task sizes only and physical resources only, respectively. Yet, many problems are still open in this area for networking researchers to solve using their expertise in handling these types of problems.

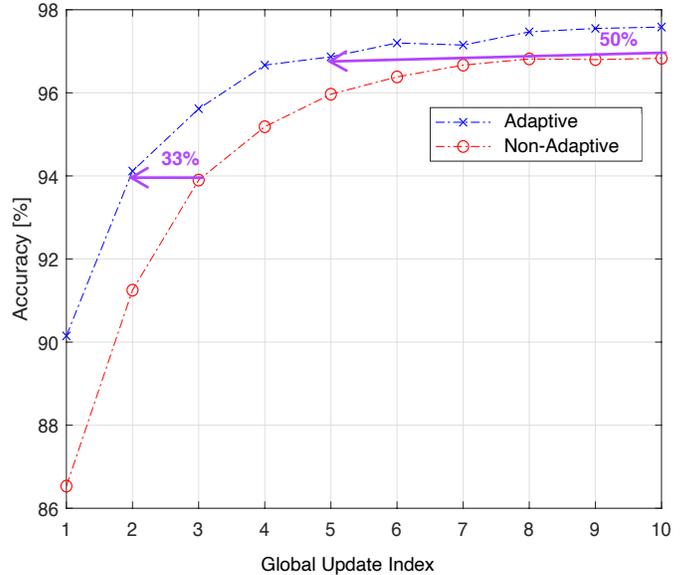

Fig. 2: Impact of task adaptation

### Mobility Considerations

The mobility of both the orchestrator and learners are two additional dimensions that, if controllable, can be jointly optimized with a subset or all the previously discussed parameters to further improve the outcomes or reduce the training time of MEL. Mobile orchestration for MEL jobs arises as a natural extension to the use of mobile servers (e.g., drone-mounted servers) in MEC. It can be of great importance for performing MEL jobs on datasets collected by a group of learners that are deployed in disaster areas (e.g., forest fires), war zones, deep mines, underwater environments, and oil/gas/water pipes. The added challenge in this setting is the determination of the optimal trajectories for these mobile orchestrators to improve the outcomes or reduce the training times of MEL jobs. The expertise in solving similar problems in MEC can be extended to solve them for the more complicated MEL settings. Further studies can then follow to jointly optimize these trajectories alongside with the task sizes and resources allocated to each of the learners.

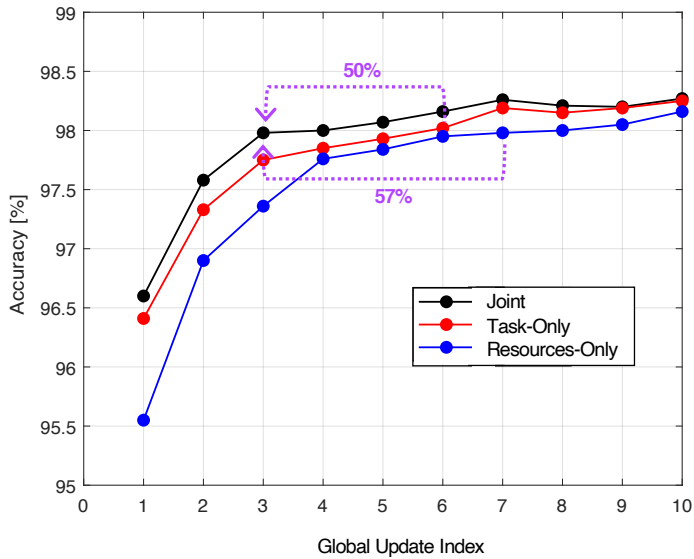

Figure 3: Impact of joint task and resource optimization

On the learners' side, mobility can be leveraged to relocate them in order to reduce the training time and/or energy. In this case, the time and energy required for such relocation are added as a factor to those consumed for the execution of the MEL job itself. The trade-off between these two components can thus be studies to identify the optimal relocation strategy minimizing the total time/energy or improving the LQL given time and/or energy constraints. Luckily, the broad knowledge in wireless node relocation for networking purposes can be leveraged as a starting point to tackle and solve these problems.

### Risk Mitigation

Several techniques can be developed to mitigate the risks of service degradation or interruption at some learners due their channel impairments, sudden reduction/ interruption in their computing service, and/or uncontrolled mobility. To avoid the degradation of the global model due to possible errors in the conveyed local parameters by some learners, [14] suggested both an elimination approach for learners having high packet error rates and a resource allocation scheme to reduce the number of such learners. Incorporating learning task redundancies can be another alternative to withstand connection losses from some learners. The new theory of coded computations [15], introduced to design coded redundancy for MEC, can be re-engineered to calculate the required redundancy for correlated MEL tasks, as a function of the number of learners and their connection/service loss probabilities. Furthermore, predictive techniques can be employed to anticipate possible connection degradation/loss at learners' parameter delivery instants and proactively mitigate such risk by changing these instants or providing more resources to these learners. Similarly, service reduction/interruptions can also be predicted and mitigated by proactively all[...] these learners or sp[...] Additionally, the m[...] can be incorporat[...] resource allocatio[...] mitigation technic[...] velocities, their p[...] and/or period(s) of[...] be predicted. These[...] and resource alloca[...] the best paramete[...] allocating tasks/res[...]

### NETWORK FORMA[...]

### Network Partition[...]

In practical FL sc[...] orchestrators requ[...] multiple MEL job[...] same network of [...] devices in the same network may need to simultaneously parallelize their learning jobs among their peers in the network. In all these settings, a partitioning of the devices into sub-networks and allocation of tasks within each of them must be performed such that each sub-network efficiently executes one of these multiple (possibly correlated) MEL jobs. The target of these processes is to achieve the best performance for all jobs given the available network resources. They must thus involve multiple physical considerations to accomplish this target, such as the channel qualities between orchestrators and learners, the available CPU power at the different learners, and their relative mobility. It also needs to take the desired LQL and temporal constraint of each job into account. Furthermore, the optimization of physical resources within each sub-network should be jointly performed with the partitioning and task allocation processes to guarantee that each sub-network can optimally execute its job thus yielding the best performance.

The joint problems of learner partitioning, task allocation, and/or resource optimization can be formulated, with different levels of details, as clustering and matching game problems within the network, as they have numerous solvers that can be executed in real-time. For uncorrelated jobs, these problems can be solved with the objective of maximizing the average of their independent LQLs (e.g., minimizing the average loss functions, average staleness, maximizing the average number(s) of local cycles, etc.). As for correlated jobs, the impact and time criticalities of these jobs can be incorporated as weight functions and/or job schedules in the partitioning, tasks allocation, and/or resource optimization decisions, to guarantee the fulfillment of the global learning target from all of them. The weight functions can be smartly designed (e.g., assigning more important jobs to more resourceful learners) to serve the purpose of this global target. Jobs can also be optimally pipelined to adjust their relative parameter delivery times so as to achieve the global learning target. Clearly, all such tools and considerations are widely used in wireless networking and can be extended to solve the aforementioned problems in MEL.

### Chain Formation for Cascaded Learning

As mentioned earlier, ML/AI models are not only continuously growing in sizes and depth, but are also concatenated to extract more features and learn from multimodal and/or multi-dimensional data. In many settings, the available learners to train these models can be either low-capacity devices that cannot bear the training of deep or concatenated models (e.g., small sensors embedded in a pipe) or more powerful devices that choose to have low energy/resource contributions to MEL jobs. The formation of chains of such devices, such that each device implements one constituent of the concatenated model or even sub-layers of this constituent can be the only solution in such settings. Fig. 4 illustrates an example of this scenario, where the constituents and sub-layers of a concatenated CNN-RNN model are split into three stages hosted by multiple chains of learners. In this formation, each learner passes its stage's outputs and local parameters to the next stage in the chain. Last-stage learners pass the accumulated local parameters of their chains to the orchestrator, which computes the global parameters from all of them at the end of each global cycle.

The problems of joint chain formation, task allocation, and/or physical resource optimization can be formulated using multi-hop cluster formation tools that are already being used for flow allocation, relay selection, and multi-hop routing. These problems will be formulated and solved given the learners' computing, inter-channel quality, and relative mobility considerations. Chain formation for multiple MEL jobs can also be investigated, and variable length chain formations for different jobs can be explored.

### Route Formation for Multi-Hop Settings

As multi-hop networking is dominant in a broad range of wireless networks, multi-hop routing tools as well as similar (even simpler) tools to those highlighted in the previous sub-section can be employed to create optimal routes for each of the learners to deliver its parameters to the orchestrator. These tools must take into account or simultaneous optimizes the learners' resources, assigned tasks, and/or positions (if mobile) so as to guarantee that all parameters from all learners are delivered on time over their multi-hop routes within the global cycle duration. Finally, parameter aggregation mechanisms such as the ones used in wireless sensor networks can be extended to aggregate the local parameters, thus reducing their forwarding delays and consumed energies.

### Conclusion

The aim of this article is to streamline the efforts of networking experts to significantly contribute to and optimize the implementation of ML/AI on networks of wireless heterogeneous devices, a new paradigm that we referred to as MEL. The review of enabling technologies demonstrates that the existing DL and MEC paradigms can be readily used to inaugurate the very preliminary versions of MEL. The challenges to further enhance and jointly optimize various types of networking and task/resource allocation procedures with the aim to achieve better MEL outcomes are then highlighted. Several representative research efforts to overcome these challenges are illustrated and categorized under two main research pillars, which are already heavily investigated in the wireless and mobile networking realm for other (simpler) purposes. Accordingly, the article has shed light on a range of possibilities to call upon and revolutionarily extend widely used problem formulation and solving tools from each of these two tracks to establish a comprehensive research and development framework for MEL.

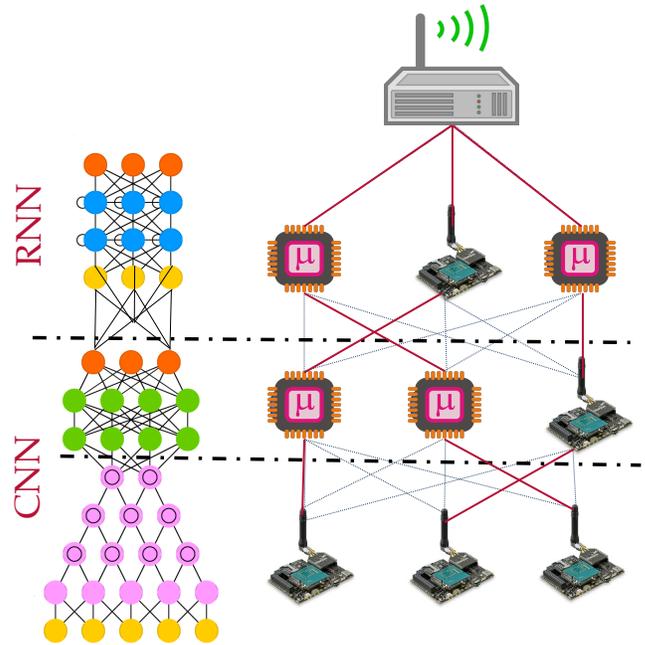

Fig. 4: An example of cascaded learning on chains of learners implementing concatenated CNN-RNN models

**Sameh Sorour [SM]** earned his Ph.D. from University of Toronto in 2011. He serves as Associate Editor in IEEE Communication Letters. His current interests include edge networking, computing, and learning for cyber-physical and autonomous systems.

**Umair Mohammad** earned his B.Sc. and M.Sc. at King Fahd University of Petroleum and Minerals in 2013 and 2016. He is currently a Ph.D. candidate at University of Idaho. His research is in the area of mobile edge learning.

**Amr Abutuleb** earned his B.Sc. from the Arab Academy for Science, Technology, and Maritime Transport in 2019 and is currently pursuing M.Sc. at Queen's University. His research is in the area of mobile edge learning.

**Hossam Hassanein [F]** earned his Ph.D. in Computing Science from University of Alberta in 1990. His research interests and contributions span the areas of wireless, mobile, sensor, and vehicular networking and computing. He is an IEEE Communications Society Distinguished Speaker and a former chair of the Technical Committee on Ad Hoc and Sensor Networks.